\documentclass[runningheads]{llncs}
\usepackage{graphicx}
\usepackage{todonotes}

\begin{document}
\title{Evaluation of Automated Image Descriptions for Visually Impaired Students \thanks{This work is financially supported by the German Federal Ministry of Education and Research (BMBF) and the European Social Fund (ESF) (Project InclusiveOCW, no. 01PE17004).}
}
\titlerunning{Evaluation of Automated Image Descriptions for VI Students}
%
\author{Anett Hoppe\inst{1}\orcidID{0000-0002-1452-9509} \and David Morris\inst{1}\orcidID{0000-0002-6347-5666} \and Ralph Ewerth\inst{1,2}\orcidID{0000-0003-0918-6297}}

\authorrunning{A. Hoppe et al.}
%
\institute{TIB---Leibniz Information Centre for Science and Technology, Hannover, Germany \and
L3S Research Center, Leibniz University Hannover, Hannover, Germany\\
\email{david.morris, anett.hoppe, ralph.ewerth@tib.eu}\\
 }
\maketitle              
\begin{abstract}
Illustrations are widely used in education, and sometimes, alternatives are not available for visually impaired students. Therefore, those students would benefit greatly from an automatic illustration description system, but only if those descriptions were complete, correct, and easily understandable using a screenreader. In this paper, we report on a study for the assessment of automated image descriptions.
We interviewed experts to establish evaluation criteria, which we then used to create an evaluation questionnaire for sighted non-expert raters, and description templates.
We used this questionnaire to evaluate the quality of descriptions which could be generated with a template-based automatic image describer. We present evidence that these templates have the potential to generate useful descriptions, and that the questionnaire identifies problems with description templates.
\end{abstract}
\section{Introduction}
Images are widely used in educational resources, 
but their usefulness is reduced for visually impaired learners.
In specialised professional settings, image descriptions are provided by experts based on their experience and accredited guidelines~\cite{reid2008,diagramGuidelines}. This is, however, not applicable to informal learning settings, especially on the Web: Open Educational Resources are available for a multitude of topics, but with only limited accessibility for students with visual impairments due to missing alternative texts (alt-texts) and image descriptions. Relying on experts to generate descriptions does not scale.

Computer vision systems have made considerable progress in recent years on topics such as image captioning \cite{park2017,shuster2019}, formula recognition \cite{zhang2018,le2019}, diagram structure identification~\cite{DBLP:conf/eccv/KembhaviSKSHF16}, and the recognition of text in scientific figures \cite{jessen2019,morris2019}. However, while a promising direction, automatic methods still do not provide results with sufficient structure and reliable completeness. In consequence, we let ourselves be inspired by prior work on accessibility technology: studies used requirements analysis to identify needs of the future users \cite{ferres2006}; and explored templates to gather image descriptions by untrained volunteers \cite{morash2015}. Other previous work suggested HTML to structure screenreader-friendly documents and used Likert scales for the evaluation of descriptions~\cite{RichRepresentations}.

From those starting points, we examine scalable ways to provide high-quality descriptions for educational image types, such as bar or pie charts. The objective is to generate these descriptions automatically based on state-of-the-art computer vision techniques. Those technologies scale, but are unlikely to provide perfect results initially. Consequently, we explore (1) a simplification of the description task by the use of structured templates derived from expert knowledge; and (2) scalable evaluation of the descriptions based on structured questionnaires.
In doing so, we setup the context necessary to use current computer vision methods for automatic description of visual educational resources.

Section \ref{sec:image_description} introduces our methodology to acquire the necessary expertise for the creation of structured descriptions and the resulting templates; the structured evaluation procedure is described in Section \ref{sec:evaluation}. 
The developed materials are made available\footnote{\url{http://go.lu-h.de/gbxfC}}.

\section{Structured image description}
\label{sec:image_description}

\subsection{User needs analysis}
\label{sec:una}
We consulted three college-educated congenitally blind people with different levels of expertise in image description:
The first expert is a researcher at a technical university. 
She trains sighted people to write image descriptions. The second expert is a teacher at a school for blind students, and has previously worked at school for the blind teaching the use of assistive technology. 
The third interviewee is a technology professional with experience using different computer environments. Our experts were selected by reference. Interview transcripts are available upon request.

The interviews served as the basis to develop guiding principles for the design of the description templates (Sec. \ref{sec:templates}) and the evaluation questionnaire (Sec \ref{sec:evaluation}):
\paragraph{Alt-text: } Alt-texts, or short descriptions should be as \textbf{concise} as possible. They should let the screenreader user know as quickly as possible whether the image is relevant for their purposes. It should further make \textbf{use of available information} -- if the author provided a title, for instance, it should be used here.
\paragraph{Long description: }\textbf{Complete and correct information} of the figure needs to be contained. Preferably, the user gets to navigate varying levels of detail. Moreover, \textbf{context} is key: Used symbols should be described with their semantic meaning, not only their shape. Similarly, \textbf{tables} need to be carefully formatted, as they are difficult to navigate. Tables need to be understandable when reading one row at a time, without referring back to the column titles.

\subsection{Description templates}
\label{sec:templates}
We drafted template structures for four commonly used types of illustrations: line and scatter plots, bar charts, node-link diagrams, and pie charts. They use an HTML format which is easily navigated with a screenreader and present the information of the image in an ascending level of granularity. They are designed to cover all information possibly contained in a plot and are thus quite detailed. The objective is to provide access to all information a seeing user might discern, but to enable a user to decide when sufficient information has been consumed. This is best exemplified by scatter plots, which might contain hundreds of data points. Instead of listing each data point (and potentially confusing the user), the area of the diagram will be grouped in sectors, and conflated information displayed for each sector. 

\section{Structured evaluation}
\label{sec:evaluation}
Besides the generation of image descriptions, their evaluation is a bottleneck. We explore the possibility of description evaluation with untrained volunteers, enabled by structured questionnaires and thus, reducing the need for expert knowledge. The questions have been developed based on the expert interviews (Sec. \ref{sec:una}), and iteratively refined in correspondence. The result is a two-stage evaluation process consisting of 15 questions in total. In the first stage, the evaluator only sees the image description and answers questions on its perceived comprehensibility and its capacity to evoke a mental image. Then, the evaluator judges the description's completeness and correctness using the displayed image.

\subsection{Experimental setup}
\label{sec:exp_setup}
We performed a comparative evaluation against a set of best-case descriptions. As a reference, we used example descriptions supplied as part of the image description guidelines developed by the National Center for Accessible Media (NCAM) and the Benetech DIAGRAM (Digital Image And Graphic Resources for Accessible Materials) center \cite{diagramGuidelines}. Nine images were used for evaluation. For each, the study participants rate the description from the guidelines, and the template-based one, allowing us to compare both scores. 

All raters evaluate two descriptions per image without knowing the respective source. Order effects are counterbalanced by randomising the sequence of evaluated descriptions. For each image, the evaluators finish the first stage of evaluation (without seeing the image) for both available descriptions, then proceed with stage two (with the image as a reference for correctness and completeness).

We report combined scores to assess our descriptions. All 15 rating items (there were three other items recording metadata) used a scale from one to five, where five was the best score possible. Thus, the maximum score was 75. For each description, we average the responses to each question, add the averages together, and report this as "description score" in Table \ref{tab:description_scores}.

Nine untrained evaluators were recruited using social media. The analysis is limited to the six of them who finished rating at least 14 of the 18 descriptions (two descriptions are rated for each of the nine example images, each is evaluated with 15 questions). 

\subsection{Results}
\label{sec:results}
\begin{table}[t]
\begin{tabular}{l|l|l|l|l}
\textbf{Image Filename} & \textbf{Image Type} & \textbf{Score (ours)} & \textbf{Score (control)} & \textbf{Difference} \\ \hline
image030                                & Bar chart                              & 64.1               & 65.2                  & -1.1             \\
image031                                & Bar chart                              & 67.1               & 65.4                  & 1.7              \\
image032                                & Bar chart                              & 72.5               & 68.1                  & 4.4              \\
image024                                & Node-link diagram                           & 54.5               & 63.8                  & -9.3             \\
image028                                & Node-link diagram                           & 57.7               & 68.3                  & -10.6            \\
Flow-Chart-1              & Node-link diagram                           & 52.9               & 65.1                  & -12.2            \\
image033                                & Line/scatter plot                       & 67.7               & 69.0                  & -1.4             \\
scatter-plot-3                          & Line/scatter plot                       & 63.6               & 55.7                  & 7.9              \\
image034                                & Pie chart                                & 67.7               & 66.0                  & 1.7             
\end{tabular}

\caption{Table of description scores.}
\label{tab:description_scores}
\vspace{-5mm}
\end{table}
Table \ref{tab:description_scores} shows the average description scores for the descriptions generated using our templates and the control examples. The first column shows the filenames of the images in the online manual \cite{diagramGuidelines}.
Four out of nine descriptions scored within two points of the control. In two cases, our descriptions even scored higher than the control. 
However, our descriptions of node-link diagrams scored worse than the controls, indicating further need to improve those.

\section{Discussion and conclusion}
\label{sec:discussion}
Online educational resources often contain images of informative nature. Due to missing or incomplete alt-texts and descriptions, those are often inaccessible for learners with visual impairments. For this reason, we have investigated the usage of structured templates to simplify the task of automatically generating high-quality descriptions and propose a procedure to evaluate the resulting descriptions \textit{without expert involvement}. The results indicate that the current structured templates successfully capture the information in simpler diagram types, such as bar and pie charts; but need further refinement for complex schemata such as node-link diagrams. While our study is a first pointer to necessary next steps, the results of the volunteer evaluation need to be complemented by an assessment involving visually impaired users to confirm the evaluation procedure. Other future work includes the addition of other image types to the template repertoire and the development of adapted computer vision methods to automatically fill the templates with diagram information. 

%
%
%
 \bibliographystyle{splncs04}
 \bibliography{iocw}
\end{document}